\documentclass[twocolumn]{aastex63}

\usepackage{xspace}

\newcommand{\orcid}[1]{\href{https://orcid.org/#1}{\textcolor[HTML]{A6CE39}{\aiOrcid}}}
\newcommand{\SNR}{\mathit{SNR}}

\newcommand\CAS{CAS\xspace}

\received{December 16, 2020}
\revised{March 31, 2021}
\accepted{May 18, 2021}
\submitjournal{AJ}

\shorttitle{Galaxy structure with deep learning}
\shortauthors{Tohill et al.}

\graphicspath{{./}{figures/}}

\begin{document}

\title{Quantifying Non-parametric Structure of High-redshift Galaxies with Deep Learning}

\correspondingauthor{C. Tohill}
\email{clar-brid.tohill@nottingham.ac.uk}

\author[0000-0003-2527-0819]{C. Tohill}
\affiliation{University of Nottingham, School of Physics \& Astronomy, Nottingham, NG7 2RD, UK}

\author{L. Ferreira}
\affiliation{University of Nottingham, School of Physics \& Astronomy, Nottingham, NG7 2RD, UK}

\author{C. J. Conselice}
\affiliation{University of Nottingham, School of Physics \& Astronomy, Nottingham, NG7 2RD, UK}
\affiliation{Jodrell Bank Centre for Astrophysics, University of Manchester, Oxford Road, Manchester UK}

\author{S. P. Bamford}
\affiliation{University of Nottingham, School of Physics \& Astronomy, Nottingham, NG7 2RD, UK}

\author{F. Ferrari}
\affiliation{Instituto de Matemática Estatística e Física,  Universidade Federal do Rio Grande, 96203-900 Rio Grande, Brasil}

\begin{abstract}

At high redshift, due to both observational limitations and the variety of galaxy morphologies in the early universe, measuring galaxy structure can be challenging. Non-parametric measurements such as the CAS system have thus 
become an important tool due to both their model-independent nature and their utility as a straightforward computational process.
Recently, convolutional neural networks (CNNs) have been shown to be adept at image analysis, and are beginning to supersede traditional measurements of visual morphology and model-based structural parameters. In this work, we take a further step by extending CNNs to measure well known non-parametric structural quantities: concentration ($C$) and asymmetry ($A$).
We train CNNs to predict $C$ and $A$ from individual images of $\sim 150,000$ galaxies at $0 < z < 7$ in the CANDELS fields, using Bayesian hyperparameter optimisation to select suitable network architectures. Our resulting networks accurately reproduce measurements compared with standard algorithms. Furthermore, using simulated images, we show that our networks are more stable than the standard algorithms at low signal-to-noise. While both approaches suffer from similar systematic biases with redshift, these remain small out to $z \sim 7$.
Once trained, measurements with our networks are $> 10^3$ times faster than previous methods. Our approach is thus able to reproduce standard measures of non-parametric morphologies and shows the potential of employing neural networks to provide superior results in substantially less time. This will be vital for making best use of the large and complex datasets provided by upcoming galaxy surveys, such as Euclid and Rubin-LSST. 
\end{abstract}

\keywords{Astronomy, Galaxies --- 
Machine Learning --- Deep Learning --- High Redshift}

\section{Introduction} \label{sec:intro}

A galaxy's morphology is a useful indicator of its assembly, interaction and star-formation history. Morphological studies have therefore proven invaluable for tracing the evolution of the galaxy population over cosmic time. However, the faintness and small angular size of galaxies at high redshift ($ z \gtrsim 1$) makes them difficult to classify in the same manner as those nearby. Cosmological dimming causes more subtle features, such as spiral arms, to rapidly disappear with increasing redshift, leaving only the brightest galaxy components detectable \citep{FERENGI}.

Furthermore, the traditional Hubble sequence is of limited applicability at high redshift. At early times, higher rates of star-formation and merging increase the prevalence of more varied and irregular morphologies \citep{Abraham_galmorf,Elmegreen,Conselice_galmorf,mortlock}. For studies of distant galaxies we need to consider more general and robust approaches to characterising galaxy structure.

Galaxy structure can be studied using both parametric and non-parametric methods. Parametric approaches fit analytic models, such as the S{\'e}rsic profile \citep{Sersic}, to a galaxy's light distribution (e.g. \citealt{Galfit,Buitrago,Simard,Haubler,Robotham}). Such parametric methods are valuable for classifying symmetrical Hubble-type galaxies. However, they break down for more irregular, peculiar-type galaxies, as they assume a smooth light distribution. Non-parametric methods make no such assumptions. They are therefore more applicable to the variety of galaxies seen in the more distant universe, such as those with `clumpy' morphologies (\citealt{Abraham94,Noguchi_Clumpy,Bershady}).

Motivated by these considerations, a number of authors, including \citet{Abraham94, Abraham96},  \citet{schade}, and \citet{Conselice97} focused on two such non-parametric parameters, the concentration (C) and asymmetry (A) of a galaxy's light distribution. It has been shown that the concentration parameter correlates with the bulge-to-disk ratio ($B/D$) of a galaxy, while the asymmetry parameter is a good indicator of the merger history of the galaxy (\citealt{Conselice,lotz08,Nevin}). Using these parameters they were able to separate galaxies into their morphological type based on their position in this $C-A$ plane. \citet{Conselice} expanded on this by introducing a third parameter, the smoothness (S) of a galaxy's light distribution, creating the \CAS system which has become one of the most common non-parametric measures of galaxy structure.
This system has since been used in many investigations of galaxy structure across a wide range of redshifts (e.g.\ \citealt{Yagi, Hoyos}). A variety of similar non-parametric statistics are also in use (e.g.\ \citealt{Lotz,MID}).

With the imminent arrival of large imaging surveys from new facilities, such as the Euclid, Rubin and Roman telescopes, it is of paramount importance to look into the efficacy of existing methods for measuring galaxy structure.  For example, parametric structural measurements are often very time-consuming to apply to large surveys. Non-parametric measurements are generally faster, but the algorithms are still typically applied to individual galaxies in series. While the problem is `embarrassingly parallel', significant computational resources are required to measure large numbers of galaxies in a timely fashion. With the future of extragalactic astronomy moving to extremely large surveys, it is useful to explore more computationally efficient approaches.

One increasingly popular technique, which has already proved useful in a number of areas of astronomy (\citealt{frontera-pons, Disanto,Pearson}), is machine learning. In particular, deep learning, utilizing neural networks, can apply sophisticated analyses to large datasets at a much faster rate than conventional methods (e.g.\ \citealt{Tuccillo}). Deep learning has been applied to the morphological classification of both nearby (\citealt{Dieleman,Sunny}) and distant galaxies \citep{Huertas_candels, Leo_merger}. It has also been shown to be very effective at reproducing parametric structural measurements \citep{Tuccillo}. However, as yet, deep learning has not been applied to determine the non-parametric CAS parameters. Given the arguments above, this 
could be a highly valuable tool for studying the local and high-redshift galaxy population in the next generation of surveys.

In this work we therefore create neural networks capable of predicting concentration and asymmetry parameters from a galaxy's image. (For now we neglect the smoothness parameter as it is more difficult to measure at high redshifts and needs a separate treatment.) We show that our networks are consistent with conventional algorithms in their output, and demonstrate reliable behaviour down to very low signal-to-noise ratios.
Furthermore, we find that our trained network is able to analyse $\sim 10,000$ galaxies in under 1.5 seconds, much faster than convention methods, making it well-suited to the large number of galaxies in future surveys.

This paper is organised as follows. In \S\ref{sec:data} we introduce the imaging data used in this work and describe how the conventional \CAS parameters are measured using the \textsc{Morfometryka} software \citep{Morfometryka}. The pre-processing of the data and all data augmentation is detailed in \S\ref{sec:pre-pro} and \ref{sec:data-aug}. In \S\ref{sec:cnns} and \ref{sec:bayes-opt} we describe the architecture and optimization of our neural networks. The resulting performance of these networks is demonstrated through a number of tests in \S\ref{sec:Results}, concluding with a brief summary in \S\ref{sec:summary}.

\section{Data} \label{sec:data}
\subsection{CANDELS Fields}

All of the images used in this project were taken with the Wide Field Camera 3 (WFC3) of the Hubble Space Telescope (HST) as part of the Cosmic Assembly Near-infrared Deep Extragalactic Legacy Survey (CANDELS). We use data from all 5 CANDELS fields: the Great Observatories Origins Deep Survey (GOODS)-North and GOODS-South fields, COSMOS, Extended Groth Strip (EGS) and Ultra-Deep Survey (UDS).

\begin{figure}
    \centering
    \includegraphics[width = \linewidth]{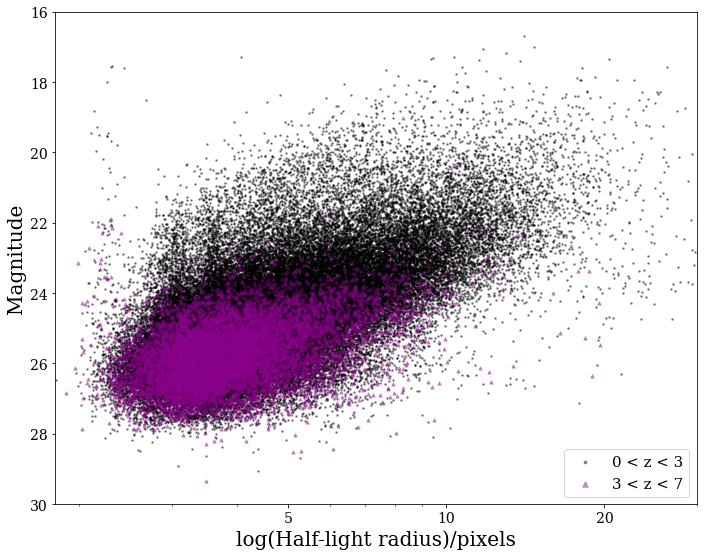}\\
    \caption{Distribution of the apparent magnitude in the H-band (F160W) vs half light radius, with a resolution of $0.06\arcsec$ per pixel, for the CANDELS galaxy sample used in this work.}
    \label{fig:mag_vs_r}
\end{figure}

The CANDELS/Deep survey ($5\sigma$ point-source limit $H = 27.7$~mag) covers an area of $\sim125~ \textrm{arcmin}^2$ with a resolution of $0.06\arcsec$ per pixel \citep{CANDELS, Koekemoer_CANDELS}. In total we have $\sim 150,000$ galaxy postage-stamp images. These galaxies have photometric redshifts covering $z = 0$--$7$, with many parameters already calculated, including star formation rates (SFR) \citep{Duncan} and \CAS values. The apparent magnitude--size distribution of our sample is shown in Fig.~\ref{fig:mag_vs_r}. In this paper we use imaging from the $H$-band ($F160$W), as it provides the most complete deep-coverage over all five CANDELS fields.

\subsection{Concentration and asymmetry}\label{sec:CAS}

As mentioned in the introduction, non-parametric methods have been used for many years to analyse the light distributions of distant galaxies, in order to better understand their structure (\citealt{Conselice,Lotz04,sazonova}). Such methods make very few assumptions, and so can be applied to peculiar and irregular galaxies as well as to classic Hubble types. 

In this paper we utilise a subset of the \CAS (Concentration, Asymmetry and Clumpiness) system as defined in \cite{Conselice}. This is a robust, non-parametric method for classifying galaxy structure, in a manner that is sensitive to their ongoing and past formation modes.  In this paper, only concentration and asymmetry are considered.  The concentration ($C$) is based on the measurement first established by \cite{Bershady}, which was found to correlate with both galaxy bulge-to-disk ratio (B/D) and the effective radius of the bulge.
This quantity is defined as
\begin{equation}
    C = 5\log_{10} \left(\frac{r_{80}}{r_{20}} \right).\label{eq:concentration}
\end{equation}
\noindent where $r_{20}$ and $r_{80}$ are the radii containing 20\% and 80\% of the total light of the galaxy, respectively. 
The value of $C$ is simply a measure of how concentrated the light in the central region is relative to the galaxy's overall size. Galaxies with higher concentrations are typically ellipticals, early-type disks, and edge-on disks. In this manner, it shares similarities with the S{\'e}rsic index \citep{Graham_sersic}.

\begin{figure}
    \centering
    \includegraphics[width = \linewidth]{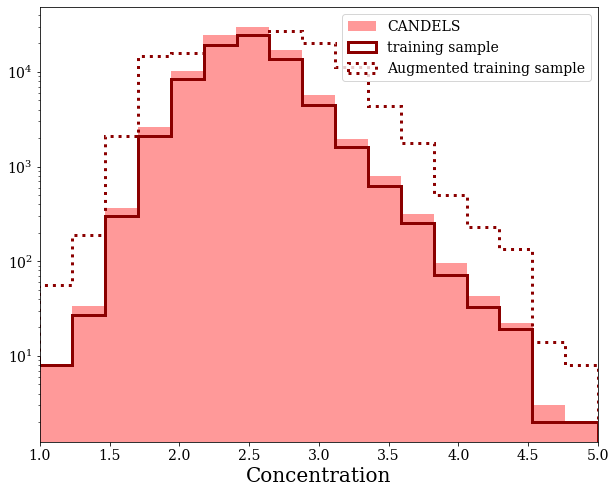}\\
    \includegraphics[width =\linewidth]{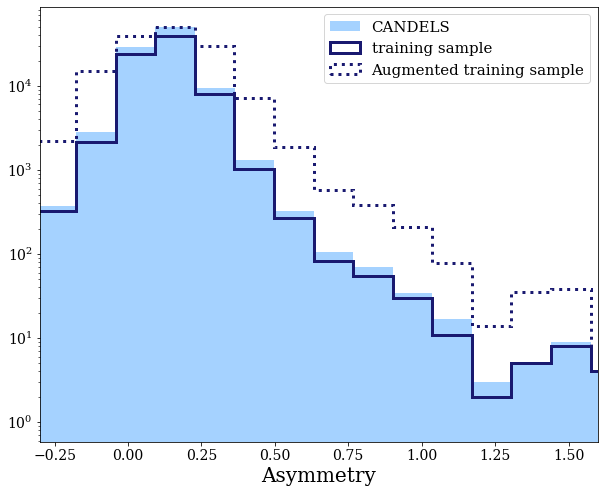}
    \caption{Distributions of asymmetry and concentration values for our selected sample of CANDELS galaxies, measured with \textsc{Morfometryka}. The solid line shows the selected training sample and the dotted line shows the training sample after our augmentation process.}
    \label{fig:AC_dist}
\end{figure}

Galaxy asymmetry was first used in a basic form by \cite{schade}, when trying to classify distant galaxies imaged with HST.
Asymmetry ($A$) is determined by rotating a galaxy $180^{\circ}$ about its center and then subtracting from the original image. The centre of rotation is determined by an iterative process that finds the minimum asymmetry. Further algorithmic details are described in \cite{Conselice2000} and \cite{Conselice}.
The absolute values of the residuals are summed and normalized by the original galaxy flux. The resulting asymmetry contains a contribution from the background noise. This is accounted for by subtracting a background term, determined by computing the asymmetry for small areas of sky near the galaxy. The basic calculation for the asymmetry is therefore given by
\begin{equation}
    A =  \frac{\sum|I - I_{180}|}{I} - A_\mathrm{bkg}\,,\label{eq:asymmetry}
\end{equation}
where $I$ is the original galaxy image, $I_{180}$ is the rotated galaxy image, and $A_\mathrm{bkg}$ is the background asymmetry (discussed further below).

Asymmetry can be used to identify a number of interesting galaxy classes, such as mergers and starburst galaxies (\citealt{Conselice97}, \citealt{Conselice2000}, \citealt{Bluck}). These types of galaxies have a higher $A$ value than regular ellipticals and disk galaxies, due to distributed areas of increased star formation. 

\subsection{Morfometryka}

\CAS measurements were originally obtained using IRAF. However, a more modern implementation, in Python, is provided by \textsc{Morfometryka}\footnote{The results in this paper are based on \textsc{Morfometryka} version 8.2} \citep{Morfometryka}. \textsc{Morfometryka} extracts a number of features from astronomical images, such as non-parametric morphology (including the \CAS parameters) and S{\'e}rsic profiles. Full details of the software can be found in \citet{Morfometryka}, however we will briefly describe how the parameters used in this paper were calculated.

\begin{figure*}
    \centering
    \includegraphics[width=\linewidth]{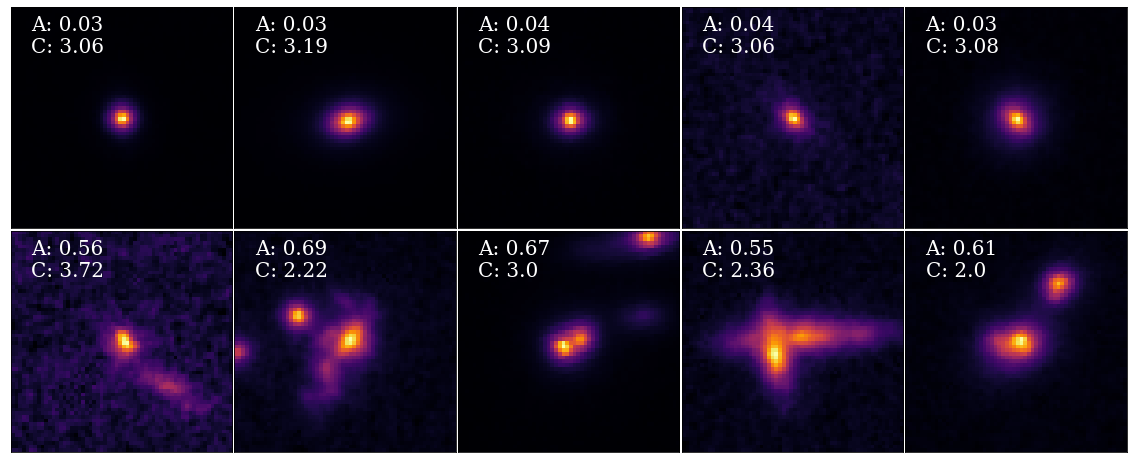}
    \caption{\emph{Top row:} Images of the galaxies with high concentration values. These galaxies appear to be compact, spheroidal and have no close neighbours. These galaxies also typically have low asymmetry values, although there are some objects with high concentration and asymmetry. \emph{Bottom row:} Images of galaxies with high asymmetry values. Many of these galaxies appear to be undergoing mergers and have tidal effects present, although there are occasional cases of line-of-sight projection. The concentration ($C$) and asymmetry ($A$) are indicated above each galaxy stamp.}
    \label{fig:C_A_imgs}
\end{figure*}

\textsc{Morfometryka} calculates the concentration, $C$, as explained in \ref{sec:CAS}, with the exception that the factor of 5 in Eq.~\ref{eq:concentration} is omitted. However, in order to remain consistent with previous studies, this factor was re-applied to our concentration values. 

The asymmetry, $A$, is also determined as described in Section \ref{sec:CAS}, by applying Eq.~\ref{eq:asymmetry} within a $1.5$ $\times$ Petrosian-radius elliptical aperture centred on the galaxy. However, the background term in Eq.~\ref{eq:asymmetry} is computed in a way that slightly deviates from the original \CAS implementation. The standard approach utilises a single background region. Originally \textsc{Morfometryka} did not include the background asymmetry correction term. For our measurements, we construct a $10 \times 10$ pixel grid over the image area outside the galaxy segmentation map. We then measure the asymmetry for each cell in the grid, according to the first term of Eq.~\ref{eq:asymmetry}. Finally, we select the median asymmetry across all the cells as our background term, $A_\mathrm{bkg}$. This ensures a robust and accurate background correction, improving upon the original background subtraction by eliminating the bias inherent in choosing only one background area. This is now incorporated into \textsc{Morfometryka}.

The errors on the concentration values are derived from those of the individual size measurements, which assume Poisson distributed fluxes. The typical error on $C$ is $\pm 0.23$. The error on the asymmetry values were calculated using the method described in \cite{Conselice}. We find that the typical error on $A$ is $\pm 0.072$ for our sample. 

We applied \textsc{Morfometryka} to all of the images in our dataset. We then select suitable galaxies for our analysis based upon the the steps described in \S\ref{sec:pre-pro}. The distributions of concentration and asymmetry values for our selected sample are shown in Fig.~\ref{fig:AC_dist}.

A subset of galaxy images were inspected to check that the measurements correspond to visual expectations. As can be seen from the top row of Fig.~\ref{fig:C_A_imgs}, galaxies with high $C$ values appear compact and spheroidal. Such galaxies typically have low $A$ values,  reflecting a broad anti-correlation between $C$ and $A$ for normal Hubble types. Galaxies with high asymmetries are shown in the bottom row of Fig.~\ref{fig:C_A_imgs}. The contrast between the two sets of galaxies is clear, with high $A$ galaxies appearing disrupted, or possessing features associated with merging, such as tidal tails and multiple bright sources. Note that high asymmetry galaxies span a range of concentrations. This is a reassuring reconfirmation of how these parameters have been seen to behave in past studies \citep{Conselice08,Conselice11}.

\section{Method}\label{sec:Method}
\subsection{Pre-Processing}\label{sec:pre-pro}

The initial images used in this analysis are $101 \times 101$ pixel cutouts, with the target galaxy in the center of each stamp. As we are only interested in training the network to predict the $A$ and $C$ values for the target galaxy, we need to remove any other sources. In order to remove neighbouring sources from the cutouts, the galclean algorithm \citep{leo_galclean} was utilised. This algorithm removes any non-central sources at a certain threshold above the background level. These masked areas are replaced with values sampled randomly from the background distribution to ensure they do not leave shapes which could be picked up by the network. 

The majority of our galaxies have a half light radius of $\sim 10$ pixels. For computational efficiency, the individual galaxy images are therefore further reduced in size to $60 \times 60$ pixels, centered on the galaxy.

Since we are interested in measuring structure irrespective of overall galaxy brightness, we individually normalize our images. The pixel values of each image are rescaled so that the maximum pixel value of each image has a value of 1. This is also a standard pre-processing procedure for deep learning. It improves learning efficiency by ensuring that the inputs to the networks are compatible with the domain of the activation functions used within the model.

We wish to consider only reliable galaxy detections, for which structural parameters can reasonably be obtained. We therefore limit our sample to galaxies above a minimum signal-to-noise. We define the average signal-to-noise per pixel for each galaxy as
\begin{equation}
    \SNR_p = \frac{L_\mathrm{tot}}{q \pi R_p^2 \sigma_{sky}}
\end{equation}
where $L_\mathrm{tot}$ is the total integrated flux within the Petrosian region (with semi-major axis $R_p$), $q$ is the axis ratio measured from the intensity distribution using the image moments, and $\sigma_{sky}$ is the standard deviation of the sky background. By visual inspection we define a selection for our galaxy sample of $\SNR_p > 2$. We also limit our sample to galaxies with $R_p > 5$ to ensure they are properly resolved.

Once these steps have been completed, we are left with 94,192 galaxy images with a median $\SNR_p \sim 4.5$. These images were then split randomly into training (80\%), testing (10\%) and validation (10\%) datasets to apply to our machine learning methods. With over 9,000 galaxies in each of our testing and validation sets, our performance estimates will be both accurate and precise.

\subsection{Data Augmentation}\label{sec:data-aug}

Unbalanced datasets, whereby there are many more galaxies at one particular value compared with others, can cause issues when dealing with both regression and classification problems in machine learning. The relative frequency of classes in the training set acts as a prior; the network may therefore be biased against identifying rare cases. In extreme circumstances, the network may fail to learn to identify rare cases at all. One way to combat this issue is by data augmentation \citep{data_aug}.

Data augmentation is primarily used as a way of creating a larger training sample, which more finely samples the space of possible inputs. It is a form of regularisation and hence helps to prevent overfitting. By selectively expanding the size of the potential training set, augmentation can also help to balance the prevalence of different classes, while still using all of the input data.

Looking at Fig.\ref{fig:AC_dist}, there is a large imbalance in the \CAS values for our sample, such that very high asymmetries are not common, nor are very low or high concentrations.
As we want our model to be accurate across all concentrations and asymmetries, we selectively apply augmentation to create a more balanced training set.   That is, we need to supplement the images that occupy the parameter space where there are few galaxies.
For the range of $C$ or $A$ values where there are around half the number of images compared to the median value, we rotated each image by $90^{\circ}$ once. Where there are relatively fewer images, we apply a greater variety of augmentations: rotating by $90^{\circ}$ 3 times and mirroring along both axes.
These images were then shuffled and added to the training set. After data augmentation, our training sample increases in size from 75,353 to 141,453 images. 

\subsection{Convolutional Neural Networks}\label{sec:cnns}
\begin{figure*}
    \centering
    \includegraphics[width = \linewidth]{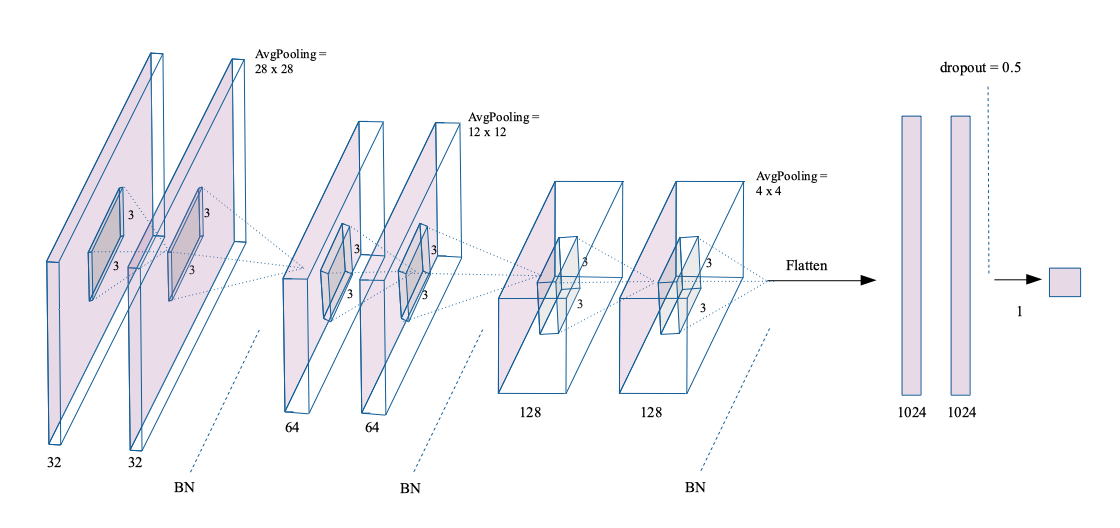}
    \caption{Architecture of the asymmetry network after optimization of its hyperparameters. This network takes input images of size $60 \times 60$ pixels, followed by 3 blocks, each containing 2 convolutional layers with 32, 64 and 128 features, respectively. Each block is followed by an average pooling layer of size 2 and a batch normalisation layer. Finally there are 2 fully-connected layers of size 1024 each, which is followed by dropout of 0.5 before the output value, i.e the network's prediction of the asymmetry.}
    \label{fig:CNN_layout}
\end{figure*}

The purpose of this project is to efficiently and robustly predict \CAS values of a galaxy from an image. We chose to implement a Convolutional Neural Network (CNN), as these are known to perform well when dealing with spatial structured data. CNNs are made up of convolutional layers, which are able to extract features from images by applying multiple filters (convolutional kernels) to the image. Individually, these filters can detect simple features. However, successive layers act hierarchically, identifying increasingly complex patterns.
One major advantage of CNNs for image classification problems is the fact that they are able to exploit the spatial structure of the data which in turn reduces the number of parameters and allows the recognition of location invariant features. 

CNNs were first popularised for image recognition/classification problems with the creation of LeNet-5 \citep{LeNet}, a network trained to classify handwritten digits. From this, CNNs have been applied in a range of fields, addressing a number of different problems and are becoming increasingly popular in astronomy. 

CNNs were first utilised for galaxy classification by \citet{Dieleman} using data from the Galaxy Zoo project \citep{Willett}. While many others had applied different machine learning (ML) techniques to address this problem (e.g., \citealt{SL92,Naim, huertas_svm,Banerji10}), these all required an earlier step of extracting features (often including CAS parameters or similar) from the images. The advent of CNNs provided a technique for efficiently extracting high-quality information directly from images. CNNs have since seen wide usage in extra-galactic astronomy, including morphological classification (e.g., \citealt{Dominguez_SDSS, Sunny, Barchi}), performing photometry (\citealt{Tuccillo, Boucaud_photomotry}), and estimating merger rates (\citealt{Leo_merger}).

There are many factors to consider when choosing the optimum architecture for a network. Many early studies based their architecture on previous studies (\citealt{Huertas_candels,Dominguez_SDSS, Aniyan_radio}), trial-and-error (\citealt{Dieleman, Feinstein_YSO}), and arbitrary choices. However, there are a number of optimisation techniques that allow these choices to be optimised in a more satisfactory manner for the problem at hand. The variety of network architectures we consider, and our method for selecting from these, are described in the following section.

To evaluate how well our networks are performing we compute the mean absolute error (MAE) and root mean squared error (RMSE) of the network's predictions. The RMSE metric also serves as our loss function. The MAE is simply a measure of the average magnitude of error between the network's prediction and the expected result,
\begin{equation}
    \mathrm{MAE} = \frac{1}{n} \sum_{i=1}^{n} |y_i - \hat{y}_i|\,,
\end{equation}
 where $n$ is the number of samples, $y_i$ is the expected value and $\hat{y}_i$ is the network's prediction. The RMSE is similar to the MAE, but it is more sensitive to large errors and so can indicate if there are many outliers present. It is calculated as
\begin{equation}
    \mathrm{RMSE} = \sqrt{\frac{1}{n} \sum_{i=1}^{n} {(y_i - \hat{y}_i)}^2}\,.
\end{equation}

\subsection{Bayesian Optimization}\label{sec:bayes-opt}

The various choices that must be made before training a network can be considered as hyperparameters. These include aspects of the network architecture, such as the number of convolutional layers and number of filters in each layer, and of the training, such as the update algorithm, learning rate and batch size. Varying these choices can significantly alter the performance of the trained network. The problem of determining which combination of hyperparameters will be best suited to a given problem typically involves a trial and error process, which is often only partially explored, or entirely neglected, resulting in a non-optimal solution.

To avoid this, many optimization techniques have been developed, from simplistic random or grid-based searches \citep{random_search}, to more advanced techniques such as random forests \citep{random_forrest}. The aim of these techniques is to find the optimum hyperparameters that will minimise the average loss. 
Traditionally, these techniques can be computationally expensive, as each variation in the hyperparameters results in a new version of the network which must be trained and then evaluated. Bayesian Optimisation \citep{Snoek} provides a more efficient solution: a record of past evaluation results are kept and used to form a probabilistic model, which the method builds upon, reducing the time to converge on a optimal model.

 Our networks comprise a number of convolutional blocks, between 1 and 3, with each block having either 1 or 2 convolutional layers. Each convolutional layer in a block has the same number of filters between 8 and 256 in powers of 2. The kernel sizes are all fixed to $3 \times 3$.
Each convolutional block is followed by a \texttt{BatchNormalisation} layer and an \texttt{AveragePooling} layer of fixed (2$ \times$ 2) size. Originally we started with \texttt{MaxPooling} layers, however we found that the networks' performance improved when using \texttt{AveragePooling} layers. (Similar behaviour was found by \citealt{Pasquet} when analysing SDSS images.)  Following the convolutional blocks we add some fully connected layers, with their number and size as hyperparameters. The number of fully connected layers ranges from 1 to 4, with each layer having the same number of filters between 128 and 1024 in powers of 2.
We include a dropout layer before our output layer as a form of regularisation, allowing the dropout rate to vary as another hyperparameter. The dropout rate is allowed to vary continuously between 0.25 and 0.60. The activation function is fixed to the common \texttt{ReLu} \citep{nair-relu} non-linearity. 

When training a network, an optimization algorithm adjusts the weights to minimise the cost function. With a plethora of optimizers now available, we have included the choice as a hyperparameter, selecting from a pool of those most commonly used, we include Adam, Adadelta, RMSprop, SGD and Adamax. We also set the learning rate as a hyperparameter, where we evaluate 5 values, 0.001, 0.005, 0.01, 0.05 and 0.1.

The parameters we defined as hyperparameters and their optimised values are displayed in Table (\ref{tab:opt_params}).

Each network was trained for a maximum of 300 epochs, but we applied "early stopping" to halt the training when the validation loss had converged, which was typically after $\sim 100$ epochs. 

Our Bayesian Optimization was carried out using the GPyOpt python package \citep{gpyopt2016}, with the aim to minimise the RMSE of the networks. Each network created during the optimization was trained and validated using the samples defined in \S\ref{sec:data}.  The MAE, RMSE and the Pearson coefficient were monitored for each iteration in the optimization. The network that had the lowest MAE and RMSE was selected as the optimum architecture for our network. 
\begin{table}
    \centering
    \begin{tabular}{|c|c|c|}
     \hline
     Hyperparameter & \multicolumn{2}{|c|}{Optimum value}  \\\cline{2-3}
      & Asymmetry & Concentration \\
     \hline
     batch size & 512 & 512 \\
     convolutional blocks & 3& 3\\
     conv.\ layers per block & 2& 2\\
     fully-connected layers & 2 & 2\\
     fully-connected layer size & 1024 & 512\\
     number of filters & 32 & 64\\
     optimization & Adamax & Adam\\
     learning rate & 0.001 & 0.001\\
     dropout & 0.50 & 0.55\\
     \hline
    \end{tabular}
    \caption{Summary of the hyperparameters selected by the Bayesian Optimization technique.}
    \label{tab:opt_params}
\end{table}

The architecture of the CNN selected for our asymmetry network is shown in Fig.\ref{fig:CNN_layout}. 
To ensure that the choice of optimum architecture is robust, we retrain multiple times, and compare the variation in the loss to the variation observed between different networks.
The variation in MAE for the optimum asymmetry network is quite stable and varies by $\sim 0.001$. Comparing the top 10 network architectures, we find that the MAE varies by $\sim 0.002$. The hyperparameters of these networks are quite similar, although the number of fully connected layers varies between 1 and 2, the dropout rate between 0.47 and 0.56, and the optimizer varies between Adamax and Adadelta. These parameters are not as significant in determining the optimum network. 

The selected concentration network has a similar architecture, with some slight variations.  
The MAE loss variation across different training runs is $\sim 0.002$, and the MAE of the top 10 architectures vary by $\sim 0.003$, very similar to above. Looking at the variation in the architectures which give equivalent performance, we see that the number of fully connected layers varies between 1 and 2 layers, the batch size between 256 and 512, the dropout rate  from 0.3 to 0.6, and the number of convolutional blocks varies between 2 and 3.

Following our use of Bayesian Optimisation and the above tests, we can be confident that our final selected networks are well-optimised. However, it is also reassuring that the performances we report below are robust to minor variations in network architecture and training.

\section{Results}\label{sec:Results}

\begin{figure}
    \centering
    \includegraphics[width = \linewidth]{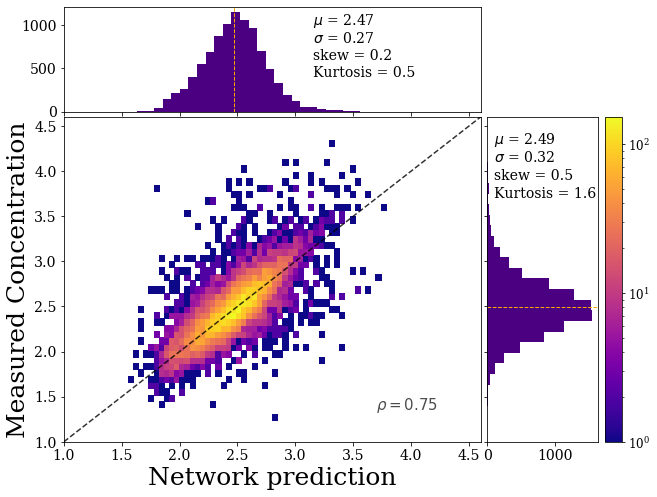}
    \caption{Our final network's predictions of concentration values for the test set versus those measured by \textsc{Morfometryka}. The network does not see any of the images used in this comparison during its training. The MAE of the network was 0.15, with a RMSE of 0.21 (see \S\ref{sec:Results}). The Pearson correlation coefficient of 0.75 indicates that there is a strong correlation between the two. }
    \label{fig:C_preds}
\end{figure}

\begin{figure}
    \centering
    \includegraphics[width=\linewidth]{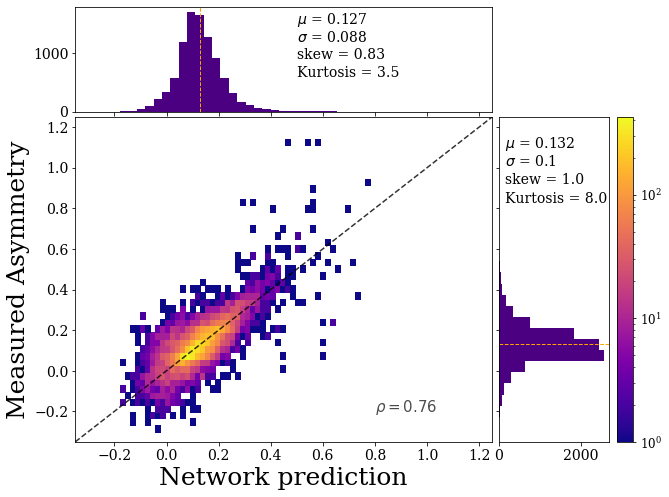}
    \caption{Our final network's predictions of the asymmetries for the test set versus those measured by \textsc{Morfometryka}. The network did not use any of these images during its training. The MAE of the network was 0.045, with a RMSE of 0.065 (see \S\ref{sec:Results}). The Pearson correlation coefficient of 0.76, indicates that there is a strong correlation between the two.}
    \label{fig:A_preds}
\end{figure}

As explained above, we train our networks on a subset (80\%) of the images and select our optimal model by its performance on a validation set (10\%). To then evaluate our selected, trained network, we use an additional independent test set (10\%; 9,420 galaxies). This ensures that the metric used to evaluate the network's performance is not biased by over-fitting the hyperparameters. We find that the networks perform similarly on both the test and validation sets: another indication that the selected network architecture is robust.

\subsection{Model performance}

The basic results of the concentration network after the hyperparameter optimisation can be seen in Fig.\ref{fig:C_preds}. The network's predictions correlate strongly with the \textsc{Morfometryka} measurements, with a MAE value of $0.15$ with a RMSE of $0.21$. This error is lower than the average error on the concentration measurements. This shows that our machine learning regression can measure these parameters just as well as the direct measurement method. Hence, the values from the network can be utilised with a similar level of confidence as the original algorithm  However, the scatter does get larger at parameter values where there are fewer galaxies.

The results for the asymmetry network, again after the hyperparameter optimisation, can be seen in Fig.\ref{fig:A_preds}. The network's predictions for the asymmetries have a MAE of 0.045 with a RMSE of 0.065. As before, this error is lower than the average error on the asymmetry measurements, showing that our networks can be used to reliably measure both the concentration and asymmetry values for a galaxy. 

Overall, both networks perform well, achieving low residuals between the measured values and the network's predictions. 
Looking at the images of galaxies where there was a large difference between our networks and the \textsc{Morfometryka}-measured \CAS values, we find that they are quite noisy, with $\SNR_p < 4$. From this we decided to further investigate the impact of noise on both our network predictions and the directly-measured \CAS values.

\subsection{Impact of noise}
\label{sec:noise}

\begin{figure}
\centering
    \includegraphics[width=\linewidth]{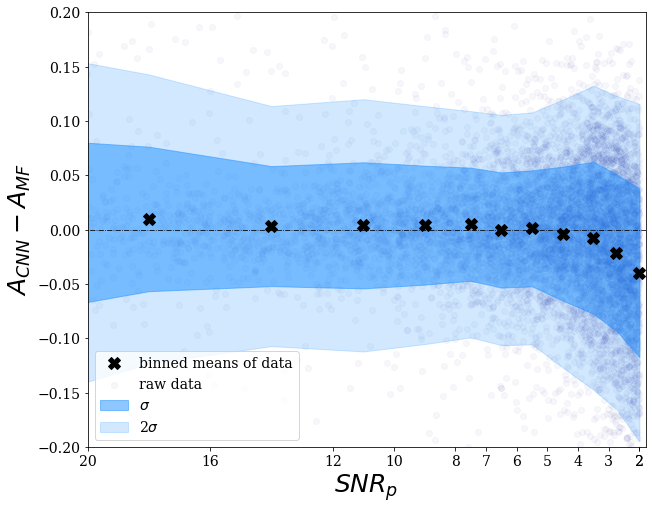}\\
    \includegraphics[width=\linewidth]{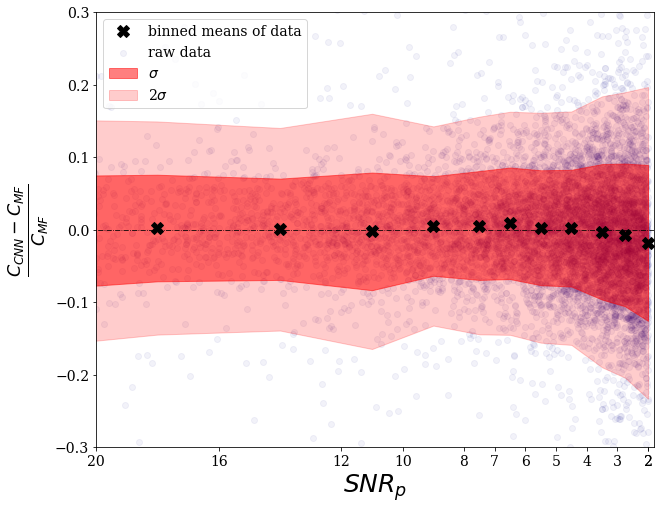}
    \caption{\emph{Top:} The residuals of the asymmetry network and the measured values from \textsc{Morfometryka} versus the $\SNR_p$ of each galaxy image. \emph{Bottom:} The fractional residuals between the concentration network and the measured values from \textsc{Morfometryka} versus the $\SNR_p$ of each galaxy image. In both panels, black points represent the means plotted for each bin with the darker shaded region representing $\pm 1$ standard deviation and the lighter shaded region shows the area containing $2\sigma$.  The trend indicates that there is a bias at low $\SNR_p$, where the networks will, on average, slightly under-predict the values measured by the standard algorithms. }
    \label{fig:SNR_vs_resids}
\end{figure}

Our networks' ability to accurately predict \CAS values is potentially dependent on the noise level in a given galaxy image. To investigate this, we consider how the residuals (network prediction $-$ \textsc{Morfometryka} value) of each network depend on the signal-to-noise per pixel, $\SNR_p$.  This is shown in Fig.\ref{fig:SNR_vs_resids}. Confirming the results from Figs.\ref{fig:C_preds} \& \ref{fig:A_preds}, we see that for galaxies with moderate and high $\SNR_p$ the residuals are close to zero. The random scatter is also fairly constant with $\SNR_p$, indicating that our networks are reliable across a broad range of $\SNR_p$.
There are a small number of galaxies with large deviations at the higher $\SNR_p$ however, when inspecting these images we find that most contain another source in the image that was not removed by the galclean algorithm. This could explain why the measurements for these galaxies from \textsc{Morfometryka} and our networks varied. Within the low $\SNR_p$ regime we find a slight bias where the networks, on average, under-predict the values measured by standard algorithms.

The origin of this systematic trend at low signal-to-noise is interesting. Our networks have been trained to reproduce the measured values, and are clearly doing so in the majority of cases. So why the deviation at low $\SNR_p$? This could be seen as a failure of our model to capture the details of the measurements. On the other hand, we apply regularisation and optimise the hyperparameters to avoid over-fitting, with the aim of producing a generally applicable model, capable of accurate measurements for a wide variety of images. One optimistic possibility is that our networks are able to learn a model which is more robust than the regular methods. This is not inconceivable, since the regular methods must make a series of algorithmic `decisions' (masking, fitting elliptical isophotes, recentering, etc.). The networks, instead, consider all of these issues within a single `holistic' calculation.

In order to determine if this low signal-to-noise trend is a bias in our networks or in the standard algorithm (as implemented in \textsc{Morfometryka}), we investigate how noise impacts these two approaches in an independent manner.   For this test, we select a sub-sample of 622 high $\SNR_p$ galaxies ($\SNR_p > 10$), with low asymmetry residuals ($|\Delta A| < 0.01$) from the validation set. These galaxies also have low residuals in their concentration values. These galaxies are chosen as both the network and \textsc{Morfometryka} predicted these galaxies to have similar parameters, and hence we can assume these to be the true values for the purpose of this test. 

\begin{figure*}
    \centering
    \includegraphics[width = \linewidth]{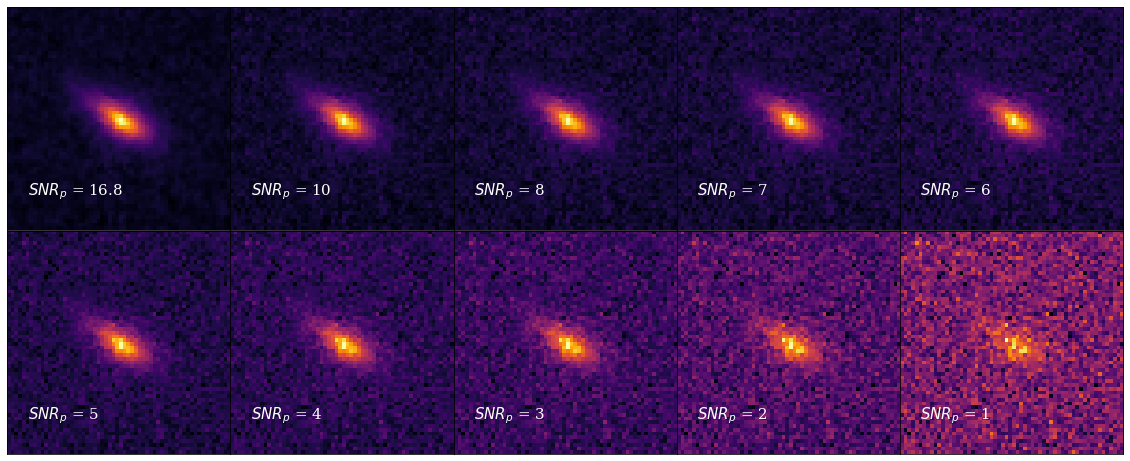}
    \caption{An example of our simulated noisy galaxy images. The top-left panel shows the original image, while the remainder show the same galaxy at different simulated $\SNR_p$.}
    \label{fig:gal_vs_SNR}
\end{figure*}

We then produce versions of each galaxy image with varying $\SNR_p$ values. To do so, we first measure the mean and standard deviation in the background of the original galaxy image, then create an image with corresponding Gaussian noise. To this simulated background image, we add the original image with the overall flux scaled, such that we achieve our desired $\SNR_p$. Finally, the image is normalized in the usual manner (\S\ref{sec:pre-pro}). Except for the variation in $\SNR_p$, each galaxy image remains identical to its original version. An example of these simulated noisy images can be seen in Fig.\ref{fig:gal_vs_SNR}.

\begin{figure}
\centering
    \includegraphics[width=\linewidth]{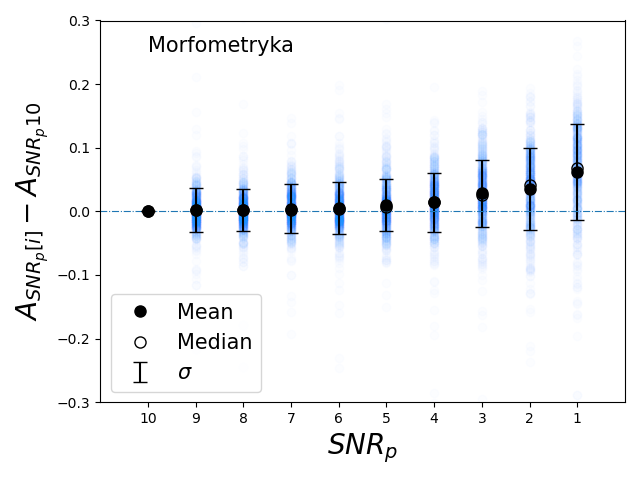}\\
    \includegraphics[width=\linewidth]{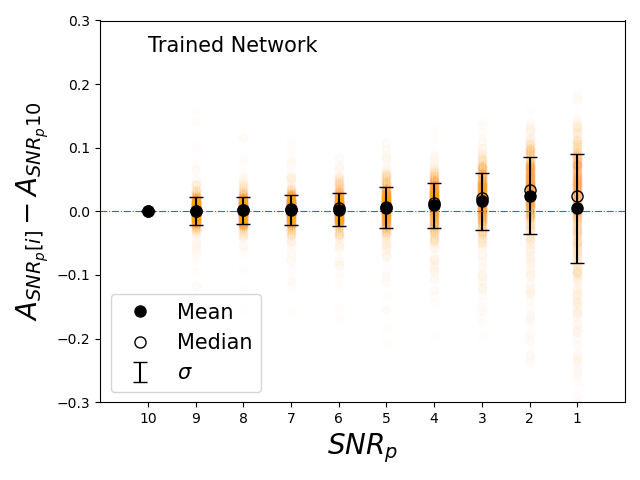}
    \caption{\emph{Top:} Deviation in the asymmetry measurements from $\SNR_p = 10$ for \textsc{Morfometryka} as the signal-to-noise is decreased in bins of $\SNR_p$. At $\SNR_p \leq 5$ there is a consistent bias, such that, on average, \textsc{Morfometryka} overestimates the values. Black points indicate the mean residual at each bin, with error-bars showing $\pm 1$ standard deviation. The median deviations are shown by open points. \emph{Bottom:} As the top panel, but for our trained network. A similar trend can be seen at low $\SNR_p$, but the scatter and systematic deviation is lower. This shows that the network is more stable and accurate than \textsc{Morfometryka} at these low $\SNR_p$ values. }
    \label{fig:SNR_vs_measurments_A}
\end{figure}

\begin{figure}
\centering
    \includegraphics[width=\linewidth]{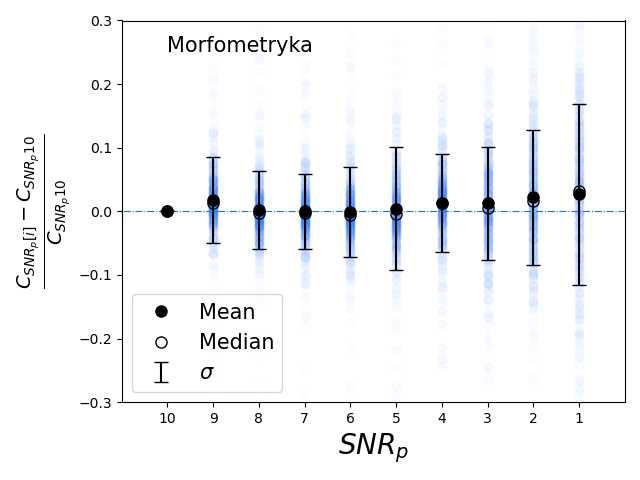}\\
    \includegraphics[width=\linewidth]{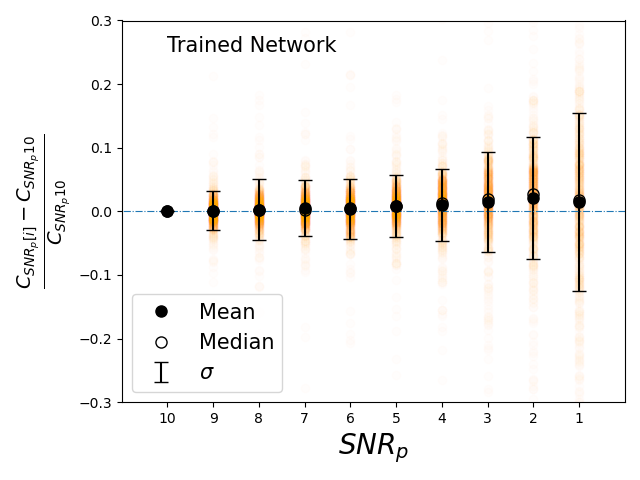}
    \caption{\emph{Top:} Deviation in the concentration measurements from $\SNR_p = 10$ for \textsc{Morfometryka} as the signal-to-noise is decreased in bins of $\SNR_p$.  Black points indicate the mean residual at each bin, with error-bars showing $\pm 1$ standard deviation. The median deviations are shown by open points. \emph{Bottom:} As the top panel, but for our trained network.  At $\SNR_p \leq 5$ there is a slight bias, such that, on average, \textsc{Morfometryka} and the network overestimate the values of the sample. It can be seen that the $C$ measurements are quite stable for $\SNR_p \geq 3$.  It can also be seen that the network produces a significantly lower scatter than \textsc{Morfometryka}}.
    \label{fig:SNR_vs_measurments_C}
\end{figure}

The asymmetry and concentration values for these galaxies are then re-measured at each $\SNR_p$, using both \textsc{Morfometryka} and our trained networks. The variation from the values measured in the $\SNR_p = 10$ image is plotted as a function of decreasing $\SNR_p$ in Figs.\ref{fig:SNR_vs_measurments_A} \& \ref{fig:SNR_vs_measurments_C}. 

For asymmetry, it can be seen that at both high and moderate $\SNR_p$ the values recovered by our network are very similar to the `true' values. For $\SNR_p > 5$ the recovered values vary with an average standard deviation of $0.025$, reflecting the uncertainties due to shot noise.

Furthermore, this scatter is significantly lower for our network than the standard algorithm. The average scatter in the asymmetry measurements, at $\SNR_p \geq 5$, is $0.025$ compared to $0.037$ for \textsc{Morfometryka}. 
Since we have already seen that our networks accurately recover \textsc{Morfometryka} measurements, this suggests that our network is using information in these moderately-noisy images that is not utilized by the \textsc{Morfometryka} algorithm.

At low  $\SNR_p$, we find a bias present in both the network and \textsc{Morfometryka}, such that the $A$ values are, on average, overestimated. However, it can be seen from Fig.\ref{fig:SNR_vs_measurments_A} that \textsc{Morfometryka} has a larger bias at these low $\SNR_p$, indicating that our network is slightly more accurate in the low $\SNR_p$ regime. At $\SNR_p = 3$ the bias in the network's asymmetry values is $0.016$ compared to $0.028$ for \textsc{Morfometryka}. 

While there is scatter in the individual measurements, on average our network is able to accurately estimate asymmetry, with little bias from the `true' value, at a lower $\SNR_p$ than the original algorithm. This is useful for merger fraction estimates, especially at high redshift, as we can now include galaxy images down to a $\SNR_p$ as low as 3 while still retrieving unbiased measurements with our network.  This means that we are able to measure reliable CAS parameters for more galaxies using deep learning and then we can with a direct measurement.

For concentration, plotted in Fig.\ref{fig:SNR_vs_measurments_C}, we again see a  a difference in the variation of the measurements for moderately-noisy images, with our network producing a significantly lower scatter than the standard algorithm. The average scatter in the concentration measurements at $\SNR_p \geq 5$ are $0.04$ for the network compared to $0.07$ for \textsc{Morfometryka}. Both the network and \textsc{Morfometryka} slightly overestimate the concentration at $\SNR_p \leq 5$. 

We also investigate the `catastrophic' fraction ($f_c$) of both \textsc{Morfometryka} and our network, that is the number of galaxies that fall outside of 2 sigma deviation from the mean. Again the network performs marginally better, with $f_c$ at $\SNR_p \geq 5$ being 4.3\% for \textsc{Morfometryka} compared to 3.3\% for our network. 

Based on these results, we conclude that our deep-learning approach is performing at least as well as traditional measurements of non-parametric structure.

\begin{figure*}
    \centering
    \includegraphics[width = \linewidth]{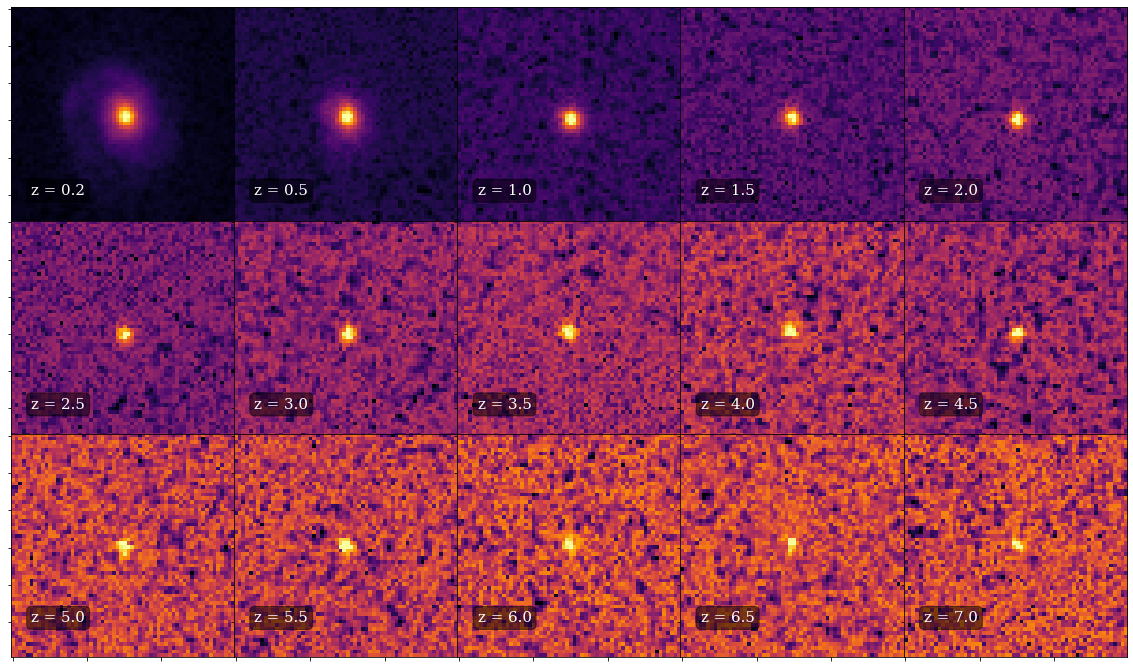}
    \caption{An example of our redshifted Frei galaxy images. In our redshifting technique we include effects due to cosmological dimming, luminosity evolution, size evolution and geometric scaling. Top left panel shows the reference image used in our tests. We start at z = 0.2 so that the whole galaxy fits within the $60 \times 60$ crop which is the input for our networks. The simulated redshift is indicated in each image. }
    \label{fig:redshifted_frei}
\end{figure*}

\begin{figure*}
    \centering
        \includegraphics[width= \columnwidth ]{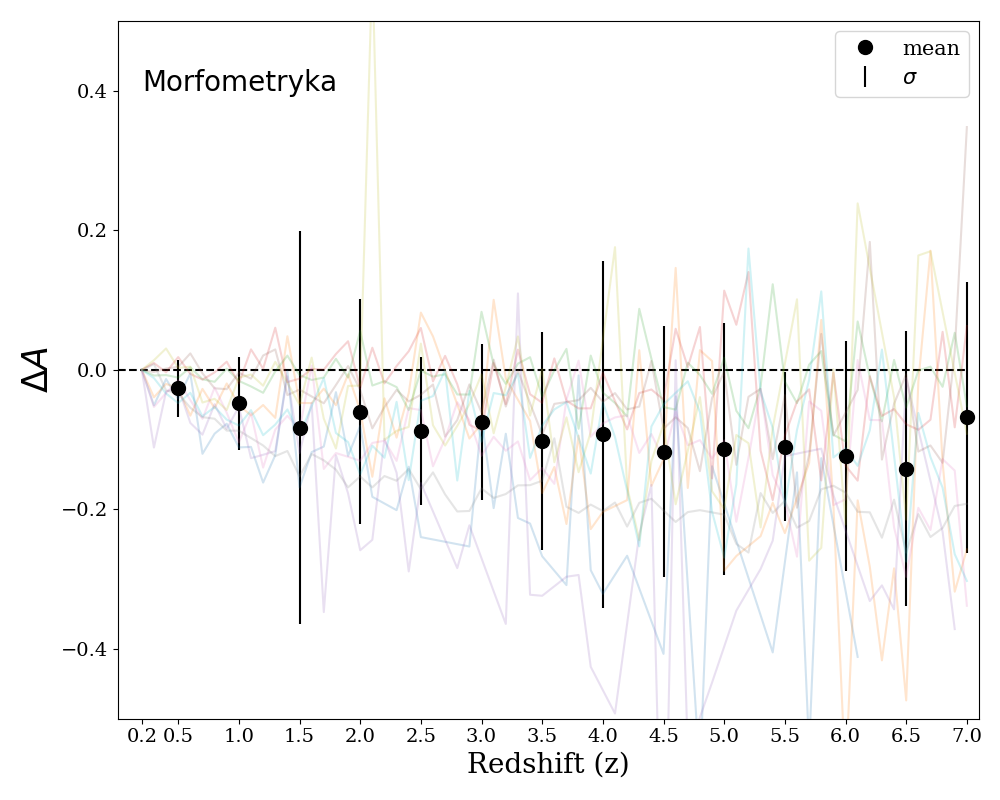}
        \hfill
        \includegraphics[width=\columnwidth]{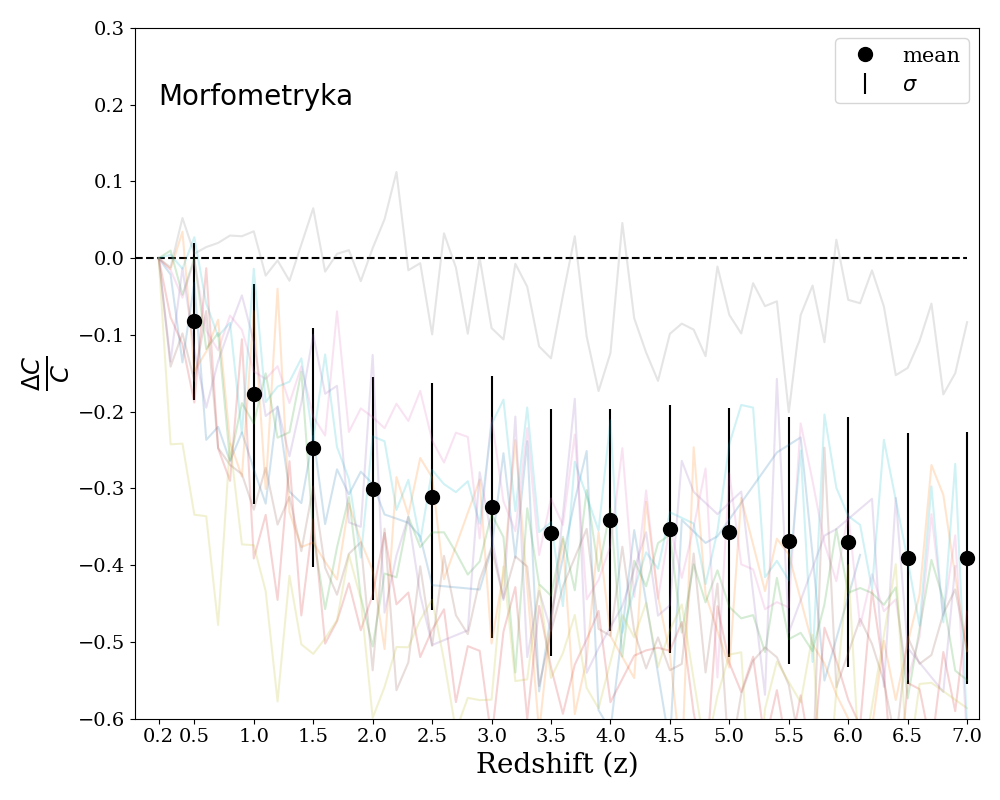}
    \\[\baselineskip]
        \includegraphics[width=\columnwidth]{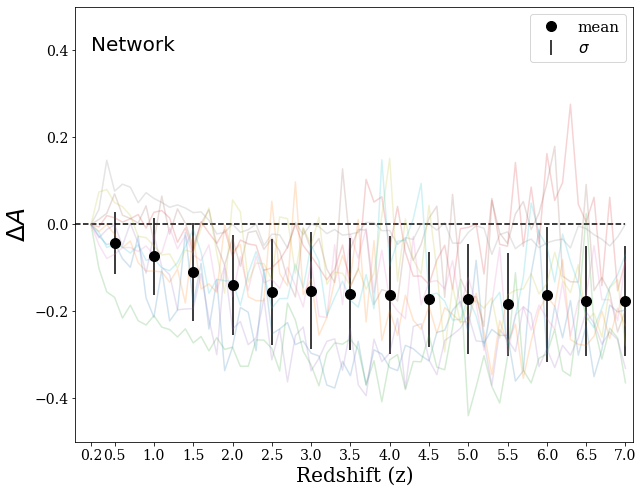}
        \hfill
        \includegraphics[width=\columnwidth]{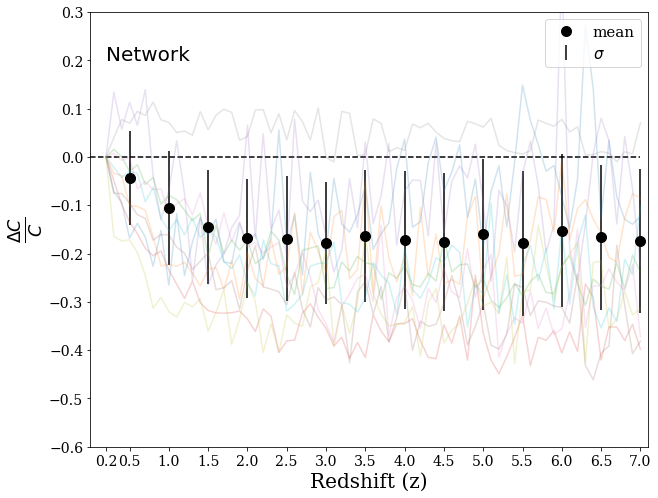}
    \caption{Variations in the $C$ and $A$ measurements from our trained networks and \textsc{Morfometryka} as a function of simulated redshift. Our simulations include geometric scaling (with accompanying reduction in resolution), size evolution, cosmological dimming and luminosity evolution. The original asymmetry measurements range from $0.01 < A < 0.48$, while the concentration measurements vary from $2.4 < C < 5.0$. There are a total of 100 galaxies from the Frei sample in this analysis. The variations for a sample of individual galaxies in the sample are shown by faint lines. These span a range of morphological types and initial asymmetry and concentration values.  Black points indicate the mean variation and the error bars show the standard deviation of the individual measurements.} 
    \label{fig:z_vs_mf_net}
\end{figure*}

\subsection{Impact of redshift effects}

While the previous test examined how our networks fare with respect to noise alone, here we combine the effects of signal-to-noise and resolution to determine our networks' performance for galaxies at high redshifts.

It is known that at higher redshifts, cosmological dimming and decreasing apparent size result in galaxies appearing more symmetric and less concentrated than they otherwise would \citep{Conselice}. However, through the use of simulations, we can model and correct for this variation.  We thus quantify the extent to which $C$ and $A$ values estimated by our networks are biased by these issues, and hence the level of any correction which should be applied when comparing galaxies at different redshifts.  

To investigate how the performance of our networks varies with redshift, we take a sample of nearby galaxies, with reliably measured concentration and asymmetry values, and simulate how they would appear at higher redshift. For this test, we selected objects from the Frei catalogue of nearby galaxies \citep{Frei}. These are a well studied sample of regular, nearby galaxies, containing all Hubble types and with previously measured $A$ and $C$ values \citep{Conselice}. We simulate the appearance of these galaxies as if they were observed at a range of redshifts, from $z = 0.1$--$7$.

There are a number of effects that need to be considered when artificially redshifting galaxies. The first effect we  address is geometric scaling, whereby the apparent size of the galaxy will decrease when viewed at a higher redshift. We follow the same procedure as described in \cite{Conselice} and \cite{galclean} to reduce the sizes of the galaxies to how they would appear at higher redshifts. Previous simulation work has kept the physical size of the galaxies constant. However, it is well known that galaxies of a given stellar mass are intrinsically smaller at higher redshifts, reducing in size by a factor of $\sim 5$ between $z = 0$ and $z = 3$ \citep{Trujillo, Buitrago}. We therefore  introduce size evolution to better represent the properties of high-redshift galaxies. 

We use the size evolution determined by \citet{Whitney}, which is based on the same $H$-band data from CANDELS GOODS North and South fields that we use in our training sample. They measured how the average physical Petrosian radius, $R_p$, changes with redshift for a mass selected sample, finding that it varies according to $R_p(z) = \frac{R_p(z=0)}{(1+z)^\beta}$, with $\beta = -0.97$. We therefore multiply the geometric scaling factor by this value to correct for the size evolution in our simulation. 

After the (flux-preserving) geometrical scaling, we apply cosmological dimming \cite{Tolman}, according to
\begin{equation}
    I(z) = \frac{I(z=0)}{(1+z)^4}
\end{equation}
where $I(z)$ is the observed intensity. This is one of the major issues when detecting high redshift galaxies, as it introduces a bias such that only the brightest, most compact galaxies are detectable. The intrinsic brightness of galaxies, with a given stellar mass, varies with redshift. We therefore implement an evolution in the surface brightness of the galaxies as outlined in \cite{Whitney_SB}. They found that the correction for the intrinsic surface brightness follows \begin{equation}
    \mu(z) = \mu(z=0)(1+z)^{\alpha} 
\end{equation} 
where $\mu = -2.5log(I)$ and  $\alpha = -0.13$. The value of $\alpha$ was found to vary from $-0.09$ to $-0.18$, but this will not result in much variation in our results. The value of -0.13 was the value found for their size corrected sample. 

To complete our simulations, the galaxies are convolved with the HST PSF in the H$_{160}$-band filter and placed in an actual CANDELS background.

We do not account for morphological or magnitude k-corrections. Instead, we test how a galaxy image would vary in restframe optical wavelengths, i.e. choosing appropriate observed filters for different redshifts.  While we are currently only able to probe restframe optical up to $z = 3$ with HST, future surveys, such as JWST, will be able to probe up to $z = 7$. Furthermore, it has been found that the \CAS parameters do not vary much between the UV and optical for star forming galaxies \citep{Conselice}. 

We select the brightest galaxies for this test, such that they are above $\SNR_p > 2$ in all images out to $z = 7$, as this was the cut off used in our training sample. 
This leaves us with 100 out of the original 112 galaxies in the Frei sample. The original asymmetry values of the sample range from $0.01 < A < 0.48$, while the concentration measurements vary from $2.4 < C < 5.0$.
The $A$ and $C$ parameters of each galaxy were  remeasured at each redshift. We only consider $z = 0.2$ onward, since we need the whole galaxy to fit within a $60 \times 60$ pixel image, for input to our networks. An example of one of the redshifted galaxies is shown in Fig.\ref{fig:redshifted_frei}.

\begin{figure*}
    \centering
        \includegraphics[width= \linewidth ]{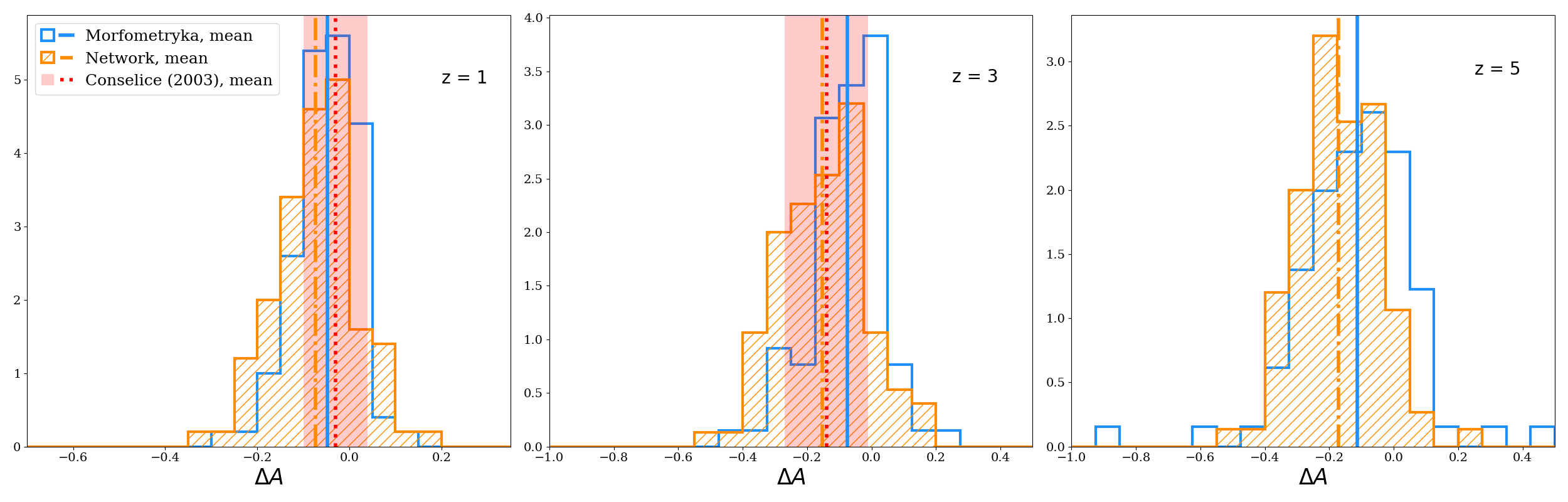}
        \hfill
        \includegraphics[width=\linewidth]{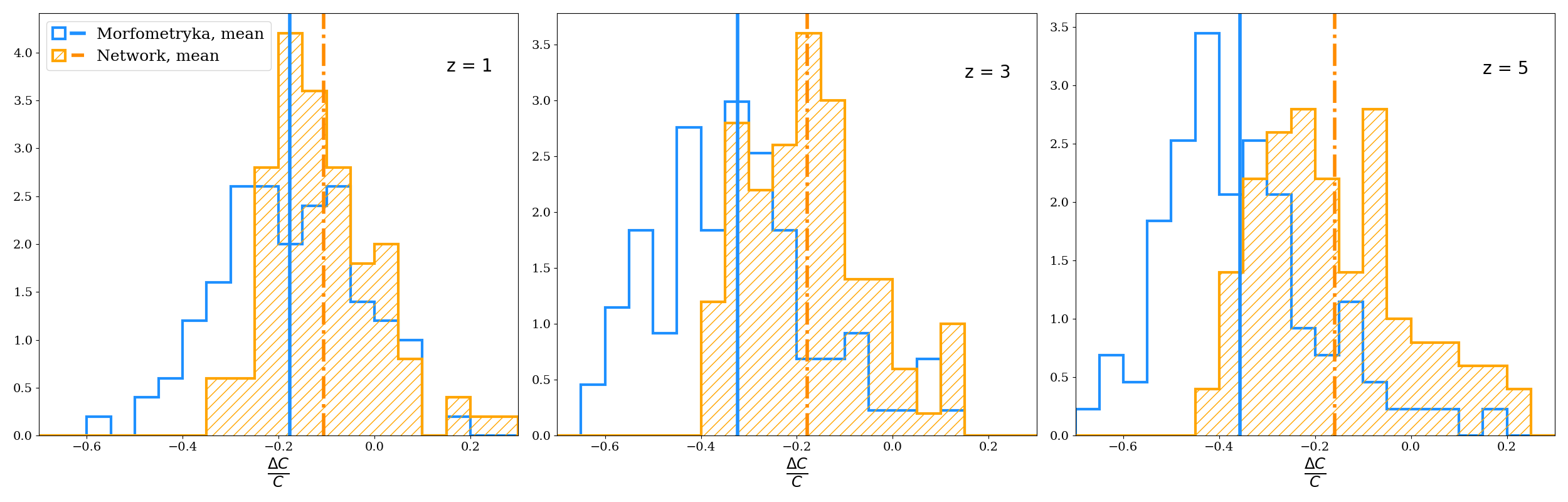}
    \caption{Investigating the effect of redshift on the asymmetry (\emph{top}) and concentration (\emph{bottom}) measurements from both our networks and \textsc{Morfometryka}. Plots show the distributions of the deviations in the measurements for a sample of redshifts. It can be seen that there are some extreme outliers in the measurements made by Morfometryka, especially with the asymmetry measurements. The range in the first asymmetry plot is reduced to better see both distributions.
    The original asymmetry measurements range from $0.01 < A < 0.48$, while the concentration measurements vary from $2.4 < C < 5.0$. There are a total of 100 galaxies from the Frei sample in this analysis. We also show the results from \cite{Conselice} (the shaded region) demonstrating that our trends agree well with those found previously.} 
    \label{fig:z_hists}
\end{figure*}

The variations of the \CAS values measured by our networks and \textsc{Morfometryka} are plotted against redshift in Fig.\ref{fig:z_vs_mf_net}, with the full distributions at a sample of redshifts shown in Fig.\ref{fig:z_hists} for comparison. As expected, at higher redshifts both methods measure the galaxies to be more symmetric and less concentrated than at $z = 0.2$. As the outer regions of a galaxy fade below the background noise, and their apparent size approaches the resolution limit of the PSF, they appear more symmetric, as has been found previously \citep{Conselice}.

While the Frei sample of galaxies appear somewhat different to those the networks were trained on, we see that the networks still perform well, measuring values similar to \textsc{Morfometryka}. The average change in asymmetry at $z = 1.0$ compared to $z = 0.2$ is $0.074$, which is similar to the average error on the high-$z$ asymmetry measurements. This is important for merger estimates, as this variation is small enough to avoid a merger appearing as a non-merger and vice versa. At redshifts higher than $z =1$ the average variation is around twice the average error on the measurements. While we cannot be sure what the equivalent $z\sim 0$ $A$ value of an individual  galaxy would be, if investigating the galaxy population at high redshift, the average $A$ value could be corrected.

We have included the equivalent results from \cite{Conselice} in Fig.\ref{fig:z_hists}. In their redshift test they included 82 galaxies from the Frei sample and investigated redshift effects up to $z = 3$. They implement no size correction in their redshifting procedure. Nevertheless, we see similar behaviour in both the bias and scatter, illustrating that the qualitative behaviour is insensitive to the details of our simulations.

These results indicate that our networks'  measurements at high redshift are better behaved than \textsc{Morfometryka}, with \textsc{Morfometryka} having a broader range of values. While both methods show systematic biases, the reduced scatter and lower prevalence of outliers suggests one could more confidently correct high-z $C$ and $A$ values based on the trained networks. 

The differences seen in this section are greater than might be anticipated purely due to $SNR$ from the results in Sec.~\ref{sec:noise}. This indicates that there are other factors affecting \textsc{Morfometryka} more than the networks. In the $SNR$ tests, simple uncorrelated Gaussian noise was added to the original images. However, in the artificial-redshifting procedure, the galaxies are placed in an apparently empty region of a real CANDELS image. The resulting background is more realistic, containing low-level structure due to pixel covariances introduced during the reduction and faint background galaxies. This could affect the asymmetry and background calculations by \textsc{Morfometryka}, especially at higher redshifts where the background subtraction becomes more significant. Our networks, which we have shown to be less susceptible to noise, appear to be less sensitive to these effects. The result is more stable measurements, which may be applied at high-redshift with greater confidence. 

Looking at the concentration measurements, we see that the network performs significantly better than the standard algorithm at recovering the original $C$ values. The network variations are, on average, a factor of two lower than those measured by \textsc{Morfometryka}. The average change in the concentration measurements between $z = 1$--$3$ is 15\% compared to 26\% for \textsc{Morfometryka}. However, as mentioned above, such systematic trends could be corrected. More importantly, the scatter in the network's measurements is somewhat lower and more consistent than those measured by the standard algorithm.

Our networks have been trained on CANDELS data, but successfully applied to data with a simple noise-degradation, and to artificially-redshifted, ground-based data. The individual images cover a very wide variety of appearances, and yet we recover reliable measurements across our test set.  This indicates that our networks are not particularly sensitive to the details of the observations. They can be applied to roughly comparable datasets with similar performance to standard methods, without the need for retraining.

The primary reason for this flexibility, is that both $C$ and $A$ measurements are determined from an image alone, without requiring any other information, such as the PSF, noise characteristics, etc. The network has learned to calculate a statistic from the image pixel values, irrespective of the observational details. It is therefore expected, but still pleasing to see, that the networks remain accurate when applied to a wide variety of images.

\subsection{Computational efficiency}

We now briefly turn to the efficiency of our CNNs, compared to \CAS measurements using conventional measurements. 
Running both our trained networks and \textsc{Morfometryka} on a single computational core, for comparison, our CNNs are able to produce measurements $\sim 3,000$ times faster. However, for a modern workstation, containing a single high-end consumer GPU (e.g. an NVIDIA GeForce GTX 1080 Ti) and 16 CPU cores, the results are even more striking.
On such a system, our trained networks can analyse $\sim 10,000$ galaxies in under 1.5 seconds, while it would take 2 hours to perform these measurements using the \textsc{Morfometryka} code.
Thus, our networks could measure all 1.5 billion resolved galaxies in the Euclid survey \citep{Euclid} on a single machine in a little over an hour. To do the same with \textsc{Morfometryka} would take several weeks on a 1000-CPU cluster! Even with highly-optimized software, using conventional algorithms would require significant time on a computing cluster.

In the previous section we have argued that our networks may be applied to other datasets without needing to retrain, providing the data characteristics are reasonably similar (which will be the case for any intermediate- to high-redshift galaxy surveys). However, should retraining be deemed necessary, this need not be an onerous process. In Sec.~\ref{sec:bayes-opt} we show that the network performance is consistent for moderate variations around the optimum. We expect that the selected hyperparameters will be a suitable choice for a variety of datasets. There should be no need to rerun the Bayesian Optimisation process again.

Given the performance we see for our networks, only a few tens of thousands of galaxies would be required to retrain the network.   
Using the optimal architecture, fully training the network with ~75,000 training examples takes only around 30 minutes. Transfer learning is also a possibility, but the training time should be no longer. In any case, the network training time is short compared to that required to prepare the training set, which itself is substantially faster than applying conventional methods to a large dataset. 

A more general argument in favour of moving towards deep learning techniques for these kinds of calculations, is that there is potential for many, currently required, preparatory steps to be avoided. Pre-processing steps such as creating segmentation maps and cleaning neighbouring objects could, in principle, be performed by the network itself. In this paper we have not explored this, and have instead applied our networks to the data prepared for Morfometryka. However, an indication of the networks' robustness is provided by its stability when applied to artificially-redshifted galaxies. This could significantly reduce the computational and human time spent preparing the data to run these measurements.

It should be noted that \textsc{Morfometryka} performs a number of additional measurements that are complementary to those discussed in this work. However, one could train a network, in the same manner presented here, to predict these parameters. Indeed, this has already been done for S{\'e}rsic profiles \citep{Tuccillo}.

A further outstanding issue is that of uncertainties. We have not attempted to produce uncertainties on individual measurements output by our networks, beyond examining the scatter relative to Morfometryka. However, \citet{Pearson21} give a detailed explanation of how the estimation of uncertainties can be incorporated into CNNs.

Extending our network to measure a wider variety of parameters, with uncertainties, is beyond the scope of this present work. However, we hope that this paper demonstrates the benefits for upcoming `Big Data' surveys, such as Euclid and Rubin-LSST. Deep-learning has the potential to improve over conventional approaches in terms of the efficiency, accuracy and flexibility with which the next generation of surveys can be analysed.

\section{Summary}\label{sec:summary}

In this paper we trained two convolutional neural networks to perform concentration ($C$) and asymmetry ($A$) measurements based on individual galaxy input images. Our trained networks reproduce measurements by standard algorithms with an average absolute error on the $C$ and $A$ values of $0.15$ and $0.045$, respectively. These are lower than the average uncertainties on those measurements using conventional methods. Our networks can therefore be used to measure these quantities with a similar level of confidence to existing algorithms.  Analysing these quantities for large samples of galaxies can provide an estimate of the merger fraction, and help us understand the transition from peculiar/irregular galaxies at high redshift to the well-defined Hubble sequence we observe locally.

We have shown how both our networks' and \textsc{Morfometryka}'s measurements are impacted by noise, but find that our networks'  estimates are more stable in the low signal-to-noise regime, in terms of both lower scatter and systematic bias. By artificially-redshifting a sample of local galaxies from the Frei catalogue, we investigate trends in the measurements due to redshift effects. 
Again, we find that our networks produce measurements with a lower level of random variation, compared to the conventional algorithms. While the measured $A$ and $C$ values are slightly biased at high-redshift, our networks and \textsc{Morfometryka} are both affected in similar manner, and consistent with behaviour seen previously \citep{Conselice}. Furthermore, the systematic offsets are comparable to the random uncertainty on individual galaxy measurements, and so relatively minor.

Our trained networks are up to several thousand times faster than previous non-parametric measurement algorithms, presenting a substantial advantage for upcoming surveys. Our trained networks are made public with this work \footnote{https://github.com/cbtohill/CASNET}.
The future of extragalactic astronomy consists of `Big Data' surveys, which will image billions of galaxies. Current state of the art computational methods for analysing these surveys will become impractical due to the computational resources and time they need. While detailed analyses will be required for certain measurements, machine learning techniques can replace many current algorithms. CNN-based approaches are more efficient and, as we have shown for measuring \CAS parameters, can be more accurate and reliable than traditional measurements.  Measuring non-parametric morphologies in upcoming galaxy surveys, including those by the Euclid, Rubin, and Roman observatories, will greatly benefit from the methods presented in this paper.  In addition, the high accuracy of our CNN-based measurements make them equally suitable for use on smaller samples from deeper surveys, such as those by JWST.

\software{Astropy \citep{astropy}, \textsc{Morfometryka} \citep{Morfometryka}, Tensorflow \citep{Tensorflow}, GPyOpt \citep{gpyopt2016}}

\acknowledgments
The authors would like to thank the Centre for Astronomy and Particle Theory of the University of Nottingham for providing the computational infrastructure needed to produce the networks used in this paper. CT acknowledges funding from the Science and Technology Facilities Council (STFC). LF acknowledges funding from Coordena\c{c}\~{a}o de Aperfei\c{c}oamento de Pessoal de N\'{i}vel Superior - Brasil (CAPES) - Finance Code 001. We thank the anonymous referee for their thorough review, which helped to significantly improve the presentation of this work.\\

\bibliography{ref}{}
\bibliographystyle{aasjournal}

\end{document}